\documentclass{sf2a-conf2018}
\usepackage{graphicx}
\usepackage{subcaption}
\usepackage{mwe}
\usepackage{hyperref}
\usepackage[]{natbib}  
\usepackage{epstopdf}

\def\BibTeX{{\rm B\kern-.05em{\sc i\kern-.025em b}\kern-.08em
    T\kern-.1667em\lower.7ex\hbox{E}\kern-.125emX}}
\bibpunct{(}{)}{;}{a}{}{,}  %%%%%%%%%%%%%  A&A bibliography style
\begin{document}

\TitreGlobal{SF2A 2018}

%%-----------------------------------------------------------------
%%      the top matter
%%

\title{Multi-scale three-dimensional visualization of emission, scattering and absorption in active galactic nuclei using Virtual Observatories tools}

\runningtitle{AGN and Virtual Observatories tools}

\author{F. Marin}\address{Universit\'e de Strasbourg, CNRS, Observatoire Astronomique de Strasbourg, UMR 7550, F-67000 Strasbourg, France}

\author{J. Desroziers$^1$}
\author{A. Schaaff$^1$}

%% Keep this line, even if the page will be settled afterwards.
\setcounter{page}{237}

%%-----------------------------------------------------------------

\maketitle

%%-----------------------------------------------------------------
%%        The abstract
%% 
%%  Warning!  within the abstract:
%%  - do not use macros. 
%%  - do not use commands like: \cite, \citet, \citep ... etc.

\begin{abstract}
Whether aimed for the study of the planetary systems, the distribution of the stars in the galaxies or the formation of the large-scale
structures in the Universe, the sizes of numerical simulations are becoming increasingly important in terms of their virtual volumes 
and computer memories. The visualization of the data becomes more complicated with the requirement of the exposition of the large number 
of data points. In order to lighten such burden, Virtual Observatories (VO) have been developed and are now essential tools in astronomy 
to share existing data, for visualization and for data analysis. Using a software, currently being developed at the Centre de Donn\'ees 
de Strasbourg (CDS) jointly with the Astronomical Observatory of Strasbourg, we show how three-dimensional radiative transfer simulations 
of active galactic nuclei (AGN) can be visualized in order to extract new information. The ability to zoom over ten orders of magnitude 
and to journey inside/between the multiple scattering regions allows to identify where emission, scattering, and absorption truly take 
place. Among all the new possibilities offered by the software, it is possible to test the single-scattering hypothesis or evaluate the 
impact of fragmentation onto the propagation of light echoes within the broad line region (BLR) or the circumnuclear region (torus).
\end{abstract}

%% Insert the keywords (to appear in the ADS indexing)
%% Keywords must be separated by a comma

%%-----------------------------------------------------------------

\section{Introduction}
%%---------------------
In the center of each massive galaxy lies a supermassive black hole \cite[see, e.g.,][]{Silk1998}, but most of those monsters 
are quiescent. Due to the lack of neighboring stars, gas and dust material they are not actively fed, which results in very 
low light emission. However, when accretion onsets and matter spirals downward the potential well, the tremendous near-infrared, 
optical and ultraviolet bolometric luminosity emitted by the system often outshines starlight emission from the host galaxy 
\citep{Pringle1972, Shakura1973}. The supermassive black hole becomes active and the object is called an active galactic nuclei 
(AGN). What is truly fascinating is that this object that has the size of a solar system can in effect have a profound impact on 
the galaxy it resides in \citep{George2018}. This involves more than ten orders of magnitudes, ranging from the Scharzschild 
radius (a few 10$^{-6}$ pc for a 10$^8$ solar masses black hole) to the extent of the polar outflows (the narrow line region, NLR) 
that can reach several kilo-parsecs. 

If we want to simulate the radiative transfer of photons from the accretion disk to a distant observer, it implies to simulate 
a variety of environments, from the highly ionized broad line region (BLR) clouds to the dusty circumnuclear torus, involving 
continuous or fragmented/filamentary structures. This demands heavy numerical calculations that are both time consuming and 
computationally expensive. Still, several softwares are now able to handle such large scale simulations 
\cite[see, e.g.,][]{Goosmann2007,Baes2011,Grosset2018}. What is less mastered, however, is the display of the results. We usually 
rely on two-dimensional projections that can suffer from projection effects such as aberrations and deformations. Three-dimensional 
visualizations are usually hampered by the large volume of data that must be loaded and stored in the computer. 

In this conference proceedings we present a new software that is currently being developed in Strasbourg as part of the global 
Virtual Observatory (VO). This tool is meant for displaying large simulations in both degraded and full resolution. The 
software allows the user to freely journey inside the simulation, isolate a given volume and create videos from several 
snapshots.

\section{3D visualization of AGN simulations}
%%-------------------------

\subsection{The radiative transfer code}
%%---------------------------------
We first create a numerical model based on the usual morphology and composition of a radio-quiet AGN. To do so, we create 
an isotropic, punctual, unpolarized radiation source at the center of the model that emits a continuous flux of 
$\lambda$ = 5510~\AA~photons. This is representative of the central supermassive black hole and its accretion disk. Around 
the disk we include a flared electron medium that flows from the torus. It's optical depth is set at unity and it's
half-opening angle from the equatorial plane is 20$^\circ$. We also include the BLR by adding 2000 ionized clouds in 
Keplerian rotation. The optical depth per cloud is also fixed at unity and the BLR sustains the same half-opening angle
than the electron flow. The volume filling factor is $\sim$ 25\%. Finally, around the BLR, we set an optically-thick, 
fragmented dusty torus with half-opening angle 30$^\circ$ from the equatorial plane. Each of its 2000 clouds has an optical
depth of 50 in the V-band. The torus onsets at the dust sublimation radius and ends around 5~pc from the center of the model. 
Finally we add a pair of collimated winds that flow from 0.01~pc to 1000~pc, using solely electrons at the wind base (optical 
thickness $\tau$ = 0.1) and a mixture of dust and electrons after the first 30~pc ($\tau$ = 0.01). The polar bi-conical 
structure has an half-opening angle of 45$^\circ$. This is a simple yet representative model of AGN based on the physical 
constraints presented in, e.g., \citet{Marin2016}.

To simulate the radiative transfer of photons in this complex and multi-scale environment we use the three-dimensional 
Monte Carlo radiative transfer code {\sc stokes}\footnote{\url{www.stokes-program.info}.} 
\citep{Goosmann2007,Marin2012,Marin2015,Rojas2018,Marin2018}. We emitted 10$^7$ photons and recorded each individual 
photon's position, polarization and timing in a binary file that represents 320~Mo (several Go in a text file).

\subsection{The visualization tool}
%%---------------------
The {\sc jasmine} application is a software created by the Centre de Donn\'ees de Strasbourg (CDS) to allow 3D visualization 
of astronomical simulations directly from web navigators \citep{Schaaff2017}. The software is essentially split into two 
parts: the client architecture, that is used for displaying data, and the server. To load a dataset in the client architecture, 
a \textit{reader} has to be written by the user and can be added directly from the interface, while the dataset itself 
must takes the form of a collection of 3D points (\textit{x},\textit{y},\textit{z}) which can possess any number of fields
(mass, velocity, temperature, polarization ...). The client can perfom various operations on the data, such as changing 
the coordinate system, displaying multiple dataset in the same window, applying filters on a given field, or creating 
animations by chaining datasets. Zooms can also be performed on specific parts of the data for more precises operations.

The navigator capacities limit the possibilities offered by the client alone, in regards of the input dataset size.
The role of the server is to bypass this limitation by allowing the user to create a database based on the simulation 
files, given a reader is created and two data-related structures are correctly filled (one for the 3D point structure 
and one for the simulation boundaries). The created database consists of two \textit{trees}\footnote{In computer science,
a tree is a widely used abstract data type that simulates a hierarchical tree structure, with a root value and subtrees 
of children (\textit{leafs}) with a parent node, represented as a set of linked nodes.}. The first tree leafs contain
the data points, and the second is a degraded representation, whose nodes and leafs consists only of averaged values 
of the 3D points contained in the first tree.

A huge (terabytes) simulation can be visualized this way by loading its whole degraded representation in the client, 
and by zooming on regions of interest. Recursive zooms can be performed on very dense areas since the loaded data are also 
represented by their degraded versions if the residuals points are too numerous. The server side implementation has been 
successfully tested on one snapshot of the 4096$^3$ particles and cells CoDa simulation \citep{Ocvirk2016}, the resulting 
databases weighting respectively 1.9~Tb (full resolution) and 510~Mb (degraded version). Note that if the client side 
can be used without the server side, the opposite is also possible by using the server-side only to perform queries on
the created database.

\subsection{Results}
%%-------------------

\begin{figure*}
    \centering
    \begin{subfigure}[b]{0.475\textwidth}
	\centering
	\includegraphics[width=\textwidth]{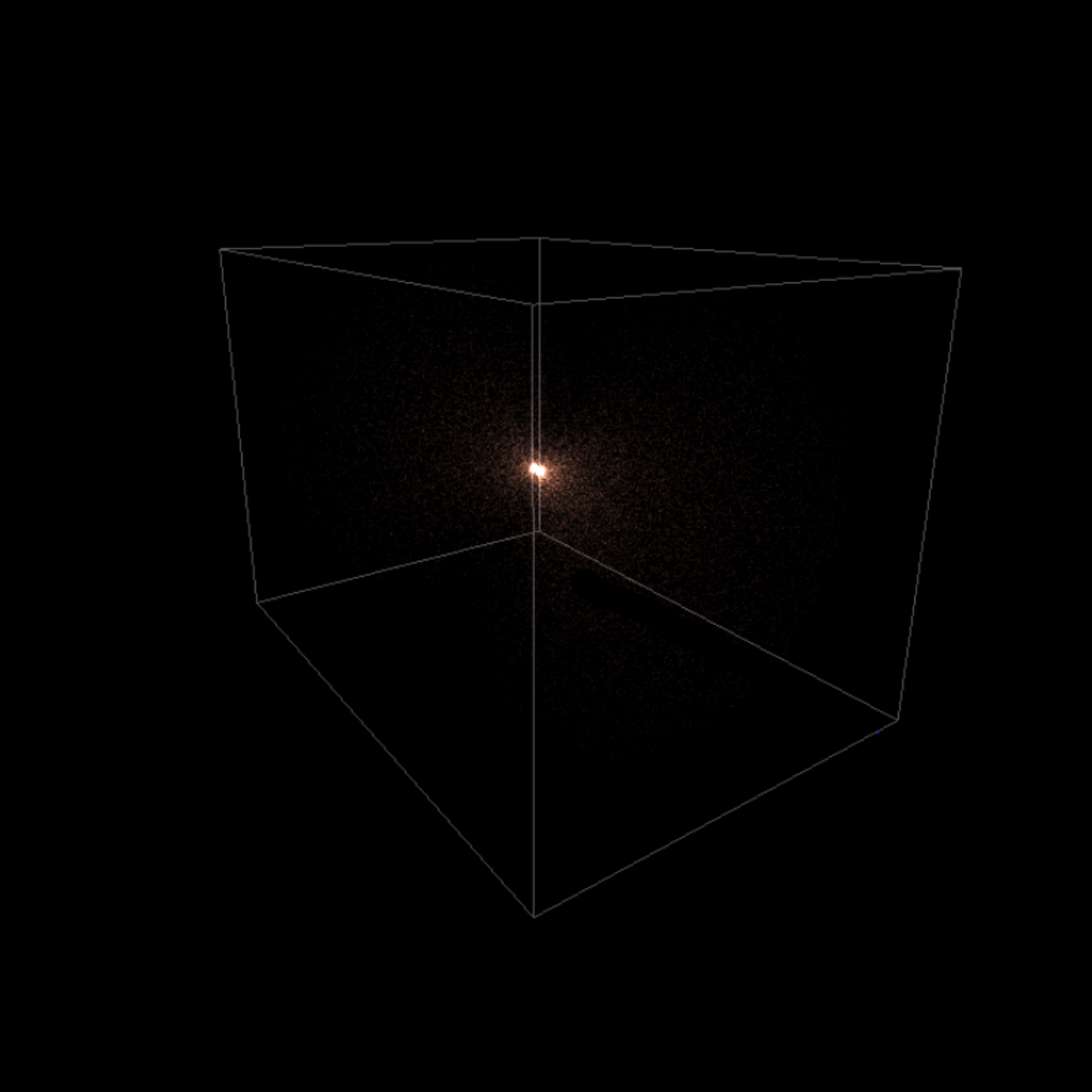}
	\caption{{\small Distance $\approx$ 50~kpc}}    
    \end{subfigure}
    \hfill
    \begin{subfigure}[b]{0.475\textwidth}  
	\centering 
	\includegraphics[width=\textwidth]{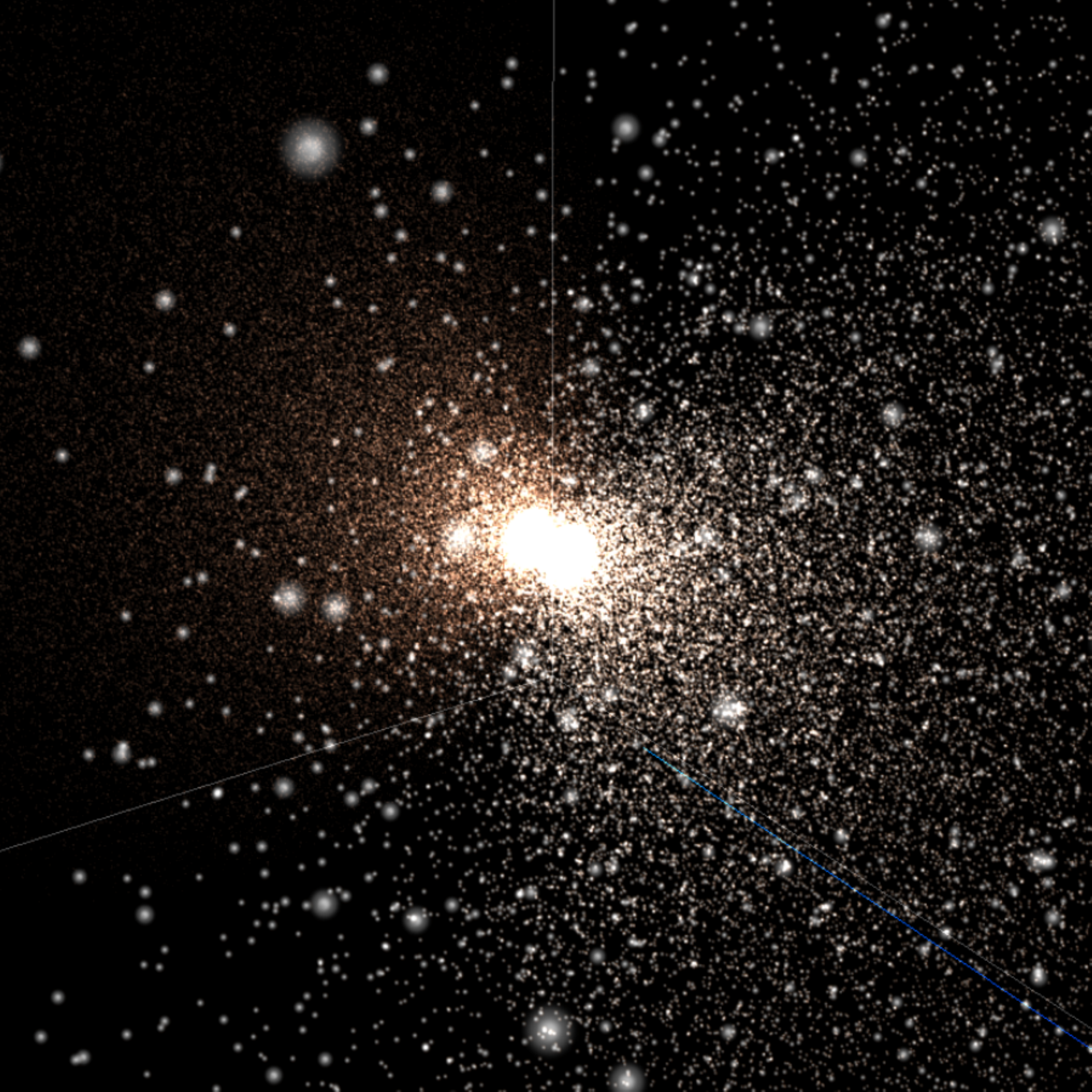}
	\caption{{\small Distance $\approx$ 500~pc}}    
    \end{subfigure}
    \vskip\baselineskip
    \begin{subfigure}[b]{0.475\textwidth}   
	\centering 
	\includegraphics[width=\textwidth]{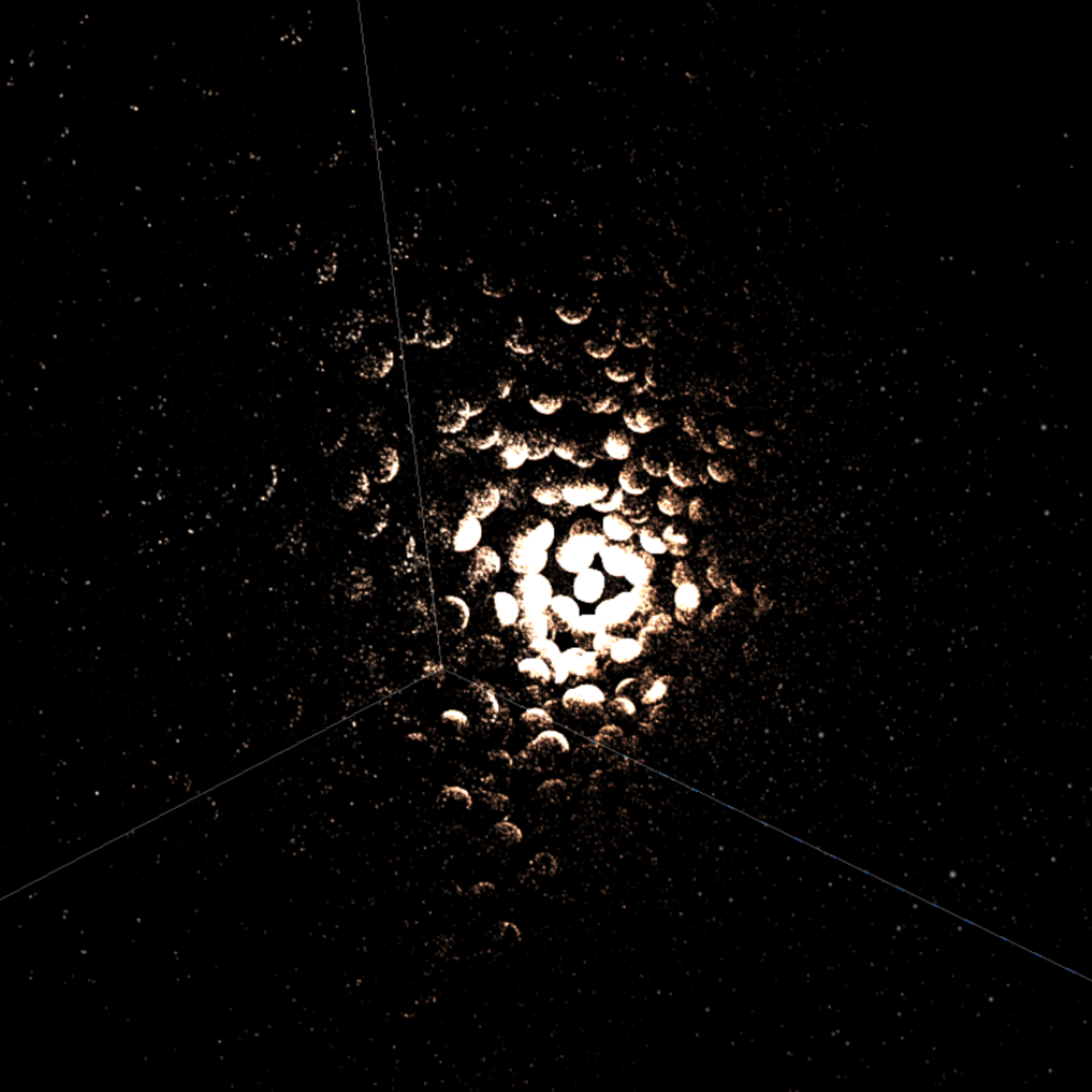}
	\caption{{\small Distance $\approx$ 500~mpc}}    
    \end{subfigure}
    \hfill
    \begin{subfigure}[b]{0.475\textwidth}   
	\centering 
	\includegraphics[width=\textwidth]{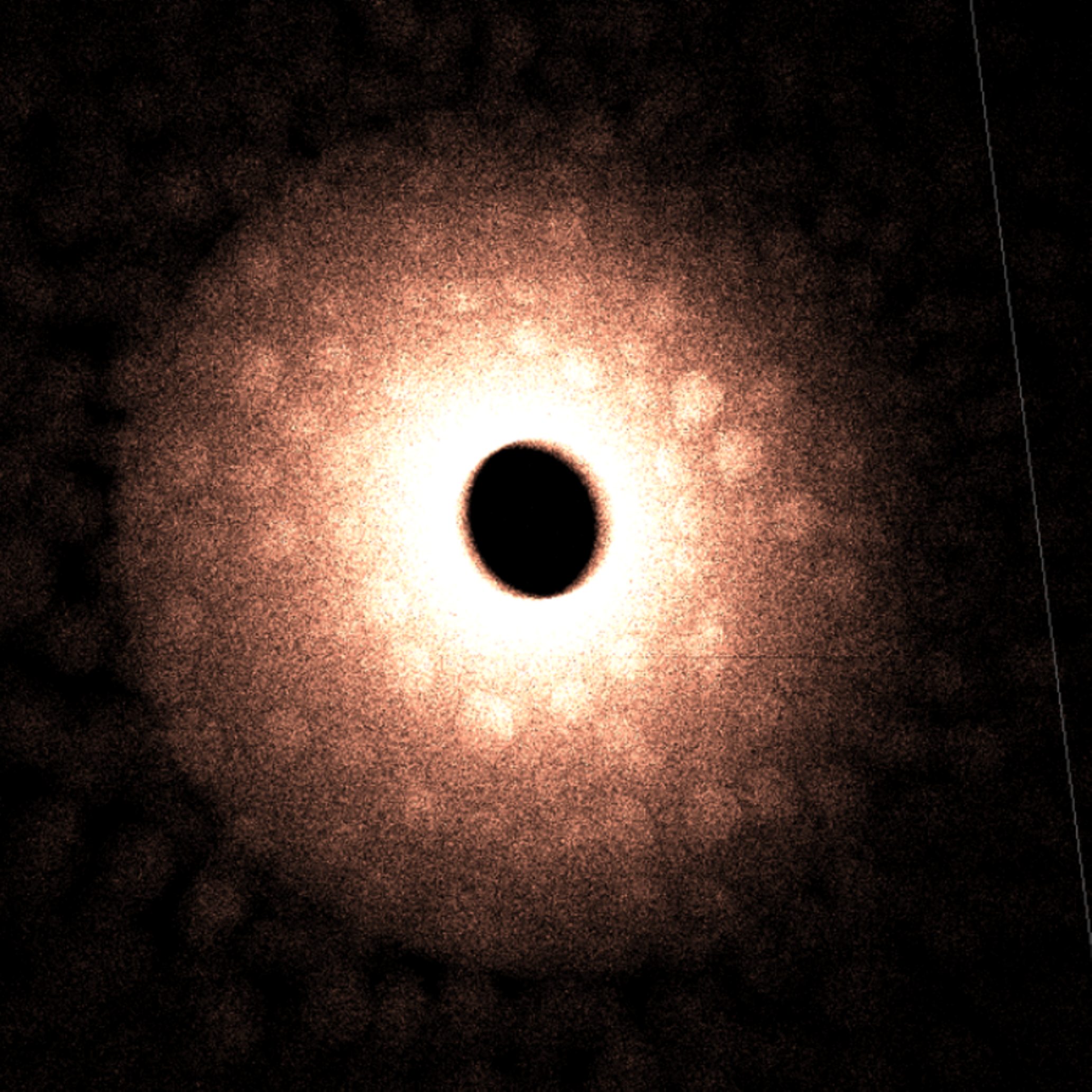}
	\caption{{\small Distance $\approx$ 50~mpc}}    
    \end{subfigure}
    \caption{Representations of the central regions of a typical 
	    Seyfert-1 galaxy, starting from 50~kpc and 
	    zooming in up to the BLR limits using 
	    a {\sc stokes} simulation and the 
	    {\sc jasmine} visualization software.
	    Each white dot marks the localization 
	    of a scattering event.}
    \label{Fig:Seyfert_1}
\end{figure*}

We show in Fig.~\ref{Fig:Seyfert_1} the results of our AGN simulation using full resolution images from the client side of {\sc jasmine}. 
We fixed the inclination of the observer to a typical type-1 viewing angle (i.e., 30$^\circ$ from the symmetry axis of the system)
that gives us a direct view of the central engine. We located the numerical camera at a distance of about 50~kpc from the center 
of the model (where the supermassive black hole resides) and highlighted in white the three-dimensional localization of each scattering
event. For this proceedings we do not show emission and absorption events but they can easily be highlighted as well. The whiter the 
dots, the more scattering events are happening. We see from Fig.~\ref{Fig:Seyfert_1} (a) that at a large distance from the AGN, 
only a point-like, quasi-stellar source in optical light is visible. Zooming in by a factor of 100 results in Fig.~\ref{Fig:Seyfert_1} (b) 
where the observer is located directly inside the polar outflowing wind. Scattering happens sporadically due to the optical thickness
of the medium ($\tau$ = 0.01) and we see that the center of the AGN clearly outshines the rest of the system. The opposite polar wind 
(whose velocity vector is opposite to the direction of the observer) appears in orange, meaning that there is less scattering events 
than in the polar outflow directed towards the observer. If we zoom in by a factor of 1000, we see in Fig.~\ref{Fig:Seyfert_1} (c)
the inner regions of the dusty circumnuclear region. The fragmented media are easily visible and the spherical clouds appear
to be brighter in the center of the model. The nest-like appearance of the AGN center \citep{Gaskell2009} creates a hierarchical 
temperature dependence of the clouds, the ones closer to the central engine being more ionized (hence warmer) than the clouds 
further away \citep{Almeyda2017}. In addition it appears that each individual cloud has a local temperature/ionization gradient due 
to self obscuration. Finally, zooming by a factor of 10 and looking at Fig.~\ref{Fig:Seyfert_1} (d), we can observe the hot flow that 
connects the central accretion disk (not shown, due to the lack of relativistic effects in the code) and the BLR clouds surrounding 
it. The central part of the flow is very bright, indicating a strong ionizing flux and many scattering events, which naturally explains
the presence of both low and highly ionized lines in the BLR \citep{Rowan1977,Osterbrock1978}. The {\sc stokes} code does not include 
strong gravity effects yet, which explains the absence of the typical relativistically warped disk at the center of the image. We detect
a gradient of scattering within the electron flow and reveal the impact of scattering onto the BLR clouds.

\section{Conclusions}
%%--------------------
In this contribution we have shown a proof of concept regarding the visualization of huge radiative transfer simulations 
using VO tools. We focused on the three-dimensional representation of scattering events in AGN using the client side of 
the {\sc jasmine} software. The numerical tool allows the observer to circulate within the simulation itself, examine where 
emission, scattering or absorption take place, and test various scenarios. For example, it is possible to simulate the disruption 
of a star by the central supermassive black hole (a tidal disruption event) and follow the resulting light echo as it propagates 
within the AGN. Several theories, such as the bird's nest appearance of the BLR, can be tested this way. The importance of multiple 
scattering is also naturally highlighted here. We intend to develop the combined use of {\sc jasmine} and {\sc stokes} in various 
situations, such as AGN variability to be probed in polarized light \citep{Rojas2018}, demonstrating the growing importance of 
VO tools in the future of astronomical visualization of large simulations.

% Optional acknowledgements
% -------------------------
\begin{acknowledgements}
This work was supported by the Centre national d'\'etudes spatiales (CNES) who funded this project through to 
the post-doctoral grant ``Probing the geometry and physics of active galactic nuclei with ultraviolet and X-ray 
polarized radiative transfer''. 
\end{acknowledgements}

\bibliographystyle{aa}  % A&A bibliography style file (aa.bst)
\bibliography{Marin} % your references in file: Yourfile.bib

\begin{thebibliography}{18}
\expandafter\ifx\csname natexlab\endcsname\relax\def\natexlab#1{#1}\fi

\bibitem[{{Almeyda} {et~al.}(2017){Almeyda}, {Robinson}, {Richmond}, {Vazquez},
  \& {Nikutta}}]{Almeyda2017}
{Almeyda}, T., {Robinson}, A., {Richmond}, M., {Vazquez}, B., \& {Nikutta}, R.
  2017, \apj, 843, 3

\bibitem[{{Baes} {et~al.}(2011){Baes}, {Verstappen}, {De Looze}, {Fritz},
  {Saftly}, {Vidal P{\'e}rez}, {Stalevski}, \& {Valcke}}]{Baes2011}
{Baes}, M., {Verstappen}, J., {De Looze}, I., {et~al.} 2011, \apjs, 196, 22

\bibitem[{{Gaskell}(2009)}]{Gaskell2009}
{Gaskell}, C.~M. 2009, \nar, 53, 140

\bibitem[{{George} {et~al.}(2018){George}, {Joseph}, {Mondal}, {Devaraj},
  {Subramaniam}, {Stalin}, {C{\^o}t{\'e}}, {Ghosh}, {Hutchings}, {Mohan},
  {Postma}, {Sankarasubramanian}, {Sreekumar}, \& {Tandon}}]{George2018}
{George}, K., {Joseph}, P., {Mondal}, C., {et~al.} 2018, \aap, 613, L9

\bibitem[{{Goosmann} \& {Gaskell}(2007)}]{Goosmann2007}
{Goosmann}, R.~W. \& {Gaskell}, C.~M. 2007, \aap, 465, 129

\bibitem[{{Grosset} {et~al.}(2018){Grosset}, {Rouan}, {Gratadour}, {Pelat},
  {Orkisz}, {Marin}, \& {Goosmann}}]{Grosset2018}
{Grosset}, L., {Rouan}, D., {Gratadour}, D., {et~al.} 2018, \aap, 612, A69

\bibitem[{{Marin}(2016)}]{Marin2016}
{Marin}, F. 2016, \mnras, 460, 3679

\bibitem[{{Marin}(2018)}]{Marin2018}
{Marin}, F. 2018, ArXiv e-prints

\bibitem[{{Marin} {et~al.}(2015){Marin}, {Goosmann}, \& {Gaskell}}]{Marin2015}
{Marin}, F., {Goosmann}, R.~W., \& {Gaskell}, C.~M. 2015, \aap, 577, A66

\bibitem[{{Marin} {et~al.}(2012){Marin}, {Goosmann}, {Gaskell}, {Porquet}, \&
  {Dov{\v c}iak}}]{Marin2012}
{Marin}, F., {Goosmann}, R.~W., {Gaskell}, C.~M., {Porquet}, D., \& {Dov{\v
  c}iak}, M. 2012, \aap, 548, A121

\bibitem[{{Ocvirk} {et~al.}(2016){Ocvirk}, {Gillet}, {Shapiro}, {Aubert},
  {Iliev}, {Teyssier}, {Yepes}, {Choi}, {Sullivan}, {Knebe}, {Gottl{\"o}ber},
  {D'Aloisio}, {Park}, {Hoffman}, \& {Stranex}}]{Ocvirk2016}
{Ocvirk}, P., {Gillet}, N., {Shapiro}, P.~R., {et~al.} 2016, \mnras, 463, 1462

\bibitem[{{Osterbrock}(1978)}]{Osterbrock1978}
{Osterbrock}, D.~E. 1978, Proceedings of the National Academy of Science, 75,
  540

\bibitem[{{Pringle} \& {Rees}(1972)}]{Pringle1972}
{Pringle}, J.~E. \& {Rees}, M.~J. 1972, \aap, 21, 1

\bibitem[{{Rojas Lobos} {et~al.}(2018){Rojas Lobos}, {Goosmann}, {Marin}, \&
  {Savi{\'c}}}]{Rojas2018}
{Rojas Lobos}, P.~A., {Goosmann}, R.~W., {Marin}, F., \& {Savi{\'c}}, D. 2018,
  \aap, 611, A39

\bibitem[{{Rowan-Robinson}(1977)}]{Rowan1977}
{Rowan-Robinson}, M. 1977, \apj, 213, 635

\bibitem[{{Schaaff} {et~al.}(2017){Schaaff}, {Deparis}, {Gillet}, {Ocvirk},
  {Steinmetz}, {Lespingal}, \& {Buecher}}]{Schaaff2017}
{Schaaff}, A., {Deparis}, N., {Gillet}, N., {et~al.} 2017, ASP Conference
  Series, 512, 503

\bibitem[{{Shakura} \& {Sunyaev}(1973)}]{Shakura1973}
{Shakura}, N.~I. \& {Sunyaev}, R.~A. 1973, \aap, 24, 337

\bibitem[{{Silk} \& {Rees}(1998)}]{Silk1998}
{Silk}, J. \& {Rees}, M.~J. 1998, \aap, 331, L1

\end{thebibliography}

\end{document}